\documentclass[useAMS,usenatbib,usegraphicx]{mn2e}
\usepackage{epsfig}
\usepackage{amsmath} 
\usepackage{rotating}           
\usepackage{color}     
\usepackage{graphicx}
\usepackage{times}
\usepackage{upgreek} 

\def \AL {$\alpha $}     

\def\lapp{\ifmmode\stackrel{<}{_{\sim}}\else$\stackrel{<}{_{\sim}}$\fi}
\def\gapp{\ifmmode\stackrel{>}{_{\sim}}\else$\stackrel{>}{_{\sim}}$\fi}

\title[QSO photometric redshifts]{QSO photometric redshifts from SDSS, WISE and GALEX colours}

\author[S. J. Curran]{S. J. Curran\thanks{Stephen.Curran@vuw.ac.nz}\\
School of Chemical and Physical Sciences, Victoria University of Wellington, PO Box 600, Wellington 6140, New Zealand}

\begin{document}

 \date{Accepted ---. Received ---; in original form ---}

\pagerange{\pageref{firstpage}--\pageref{lastpage}} \pubyear{2020}

\maketitle
\label{firstpage}
\begin{abstract}
Machine learning techniques, specifically the $k$-nearest neighbour algorithm applied to optical band colours,
have had some success in predicting photometric redshifts of quasi-stellar objects (QSOs): Although the 
mean of differences between the spectroscopic and photometric redshifts, $\Delta z$, is close to zero, the 
distribution of these differences remains wide and distinctly non-Gaussian.  As per 
our previous empirical estimate of photometric redshifts, we find that the predictions can be
significantly improved by adding colours from other wavebands, namely the near-infrared and
ultraviolet. Self-testing this, by using half of the 33\,643  strong QSO sample to train the algorithm, 
results in a significantly narrower spread in $\Delta z$ for the remaining half of the sample.
Using the whole QSO sample to train the algorithm, the same set of magnitudes return a similar spread in $\Delta z$ for a
sample of radio sources (quasars). Although the matching coincidence is relatively low (739 of the 3663 sources having
photometry in the relevant bands), this is still significantly larger than from the empirical method (2\%) and thus may
provide a method with which to obtain redshifts for the vast number of continuum radio sources expected to be detected
with the next generation of large radio telescopes.
\end{abstract}  

\section{Introduction} 
\label{intro}

There is currently much interest in developing reliable photometry-based redshifts for distant active galactic nuclei
(\citealt{lnp18} and references therein). 
Much of this is driven by the large number of sources expected to be detected through continuum surveys with the next
generation of telescopes, such as the {\em Australian Square Kilometre Array Pathfinder} (ASKAP, \citealt{jtb+08}), the
{\em LOw-Frequency ARray} (LOFAR, \citealt{vwg+13}) and the {\em extended R\"{o}ntgen Survey with an Imaging Telescope
  Array} (eROSITA, \citealt{sal14}).\footnote{See, for example, \citet{asu+17} for photometric redshift estimates of
  X-ray selected samples.}  For example, the {\em Evolutionary Map of the Universe} (EMU, \citealt{nha+11}) on the ASKAP
is expected to yield 70 million radio sources.  Being able to add a third coordinate, even statistically, would
significantly increase the scientific value of these surveys.  Obtaining the spectroscopic redshift ($z_{\rm spec}$) for
each source would be impractical thus the need for ``quick and easy'' photometric redshifts ($z_{\rm
  phot}$). Furthermore, extinction by intervening dust can make optical spectroscopy difficult \citep{cwm+06},
whereas high redshift sources,  sufficiently luminous to yield a reliable spectroscopic redshift, ionise all of the
neutral gas within the host galaxy \citep{cw12,chj+19}, biasing against the detection of gas-rich objects \citep{msc+15}.

Ideally, the redshifts for radio sources would be obtained from the radio photometric properties, although this has
proven to be elusive \citep{maj15,nsl+19}, due to their relatively featureless spectral energy
distributions (SEDs). There has, however, been considerable success applying machine learning methods of optical-band
magnitudes, specifically with the $k$-nearest neighbour (kNN) algorithm which compares the Euclidean distance between a
datum and its $k$ nearest neighbours in a feature space \citep{bbm+08}, typically the $u- g$, $g - r$, $r - i$ and $ i -
z$ colours of the {\em Sloan Digital Sky Survey} (SDSS).\footnote{There are numerous other methods used to obtain the
  photometric redshifts, for example template fitting of the SEDs (e.g. \citealt{dbw+18}). See \citet{sih18} for an
  overview.}  However, as noted by \citet{cm19}, this method fails at $z\gapp2$, giving a non-Gaussian (fat-tailed)
distribution of $\Delta z \equiv z_{\rm spec} - z_{\rm phot}$ \citep{rws+01,wrs+04,mhp+12,hdzz16}.

Finding an empirical relationship between the ratio of two colours and the redshift, \citeauthor{cm19} obtain a
near-Gaussian distribution of $\Delta z$, via a {\em redshift dependent} colour ratio, the approximate redshift first
being estimated from a near-infrared magnitude (see also \citealt{wwf+16,gas+18} and references therein).  Thus,
different combinations of observed-frame colours are required in order to yield a useful photometric redshift.  We
therefore suspect that the breakdown in the kNN method is due to the exclusive use of optical (SDSS) photometry and find
that the redshift predictions can be significantly improved with the addition of photometry from other bands, which we
address in this letter.

\section{Analysis and results}

\subsection{The sample}
\label{sec:samp}

From the SDSS Data Release 12 (DR12, \citealt{aaa+15}), we extracted the first 33\,643 
QSOs with accurate spectroscopic redshifts ($\delta z/z<0.01$), which span the magnitude range $r = 14.667 -  22.618$.  We then used the source coordinates to obtain the
nearest source within a 6 arc-second search radius in the {\em NASA/IPAC Extragalactic Database} (NED), which usually
resulted in a single match.  As well as obtaining the specific flux densities, we used the NED names to query the {\em
  Wide-Field Infrared Survey Explorer} (WISE, \citealt{wem+10}) and the {\em Two Micron All Sky Survey} (2MASS,
\citealt{scs+06}) databases.  For each of the bands\footnote{$u$ ($\lambda = 354$~nm), $g$ ($478$~nm), $r$ ($623$~nm),
  $i$ ($763$~nm), $z$ ($913$~nm) and $W1$ ($3.39$~$\mu$m), $W2$ ($4.65$~$\mu$m), $W3$ ($11.2$~$\mu$m), $W4$
  ($22.8$~$\mu$m, \citealt{bjc14}).}, 
the photometric points which fell within $\Delta\log_{10}\nu=\pm0.05$ of the central frequency of
the band were averaged, with this then being converted to a magnitude.

\subsection{QSO colours and photometric redshifts}
\subsubsection{SDSS colours}

\begin{figure*}
\setlength{\lineskip}{10pt}
\centering \includegraphics[angle=-90,scale=0.38]{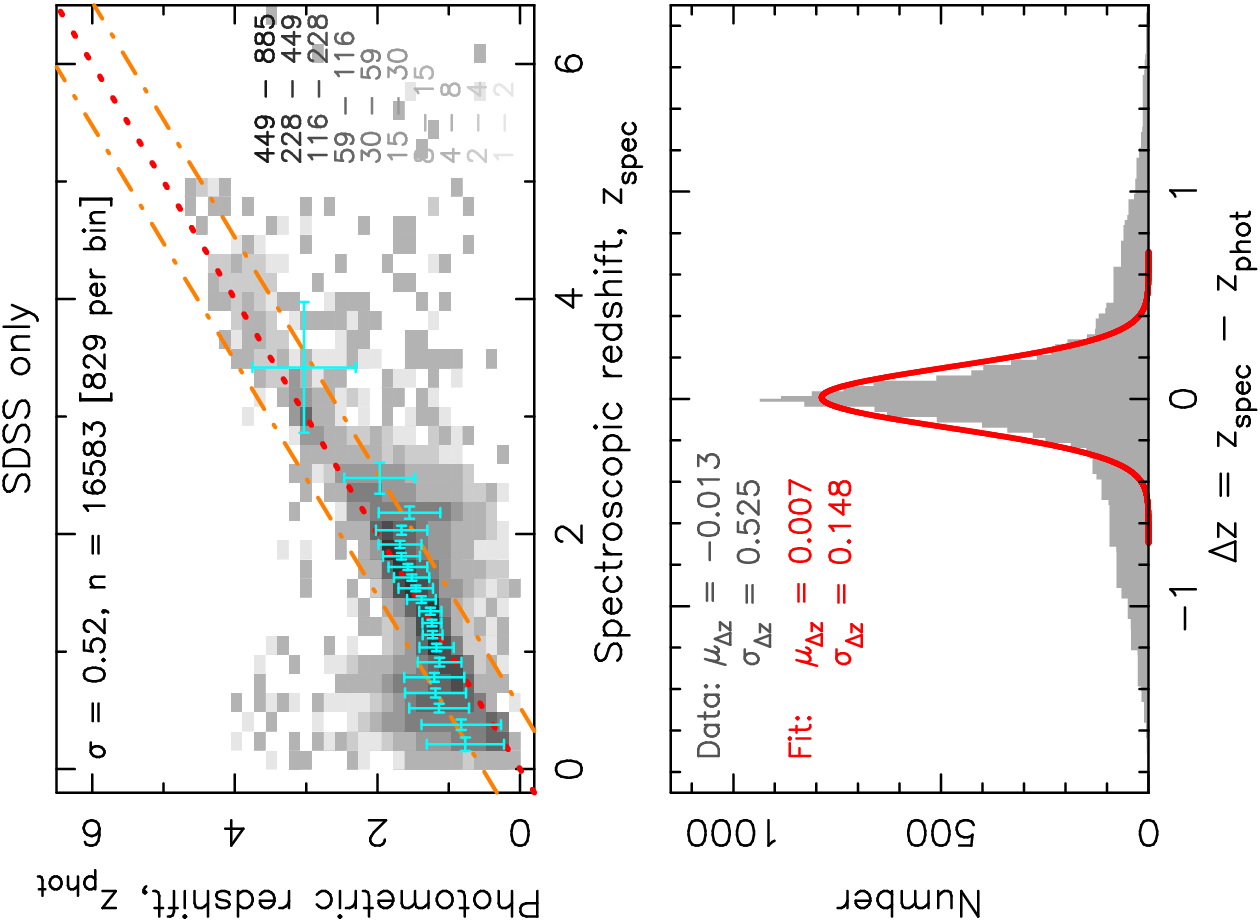}
\centering \includegraphics[angle=-90,scale=0.38]{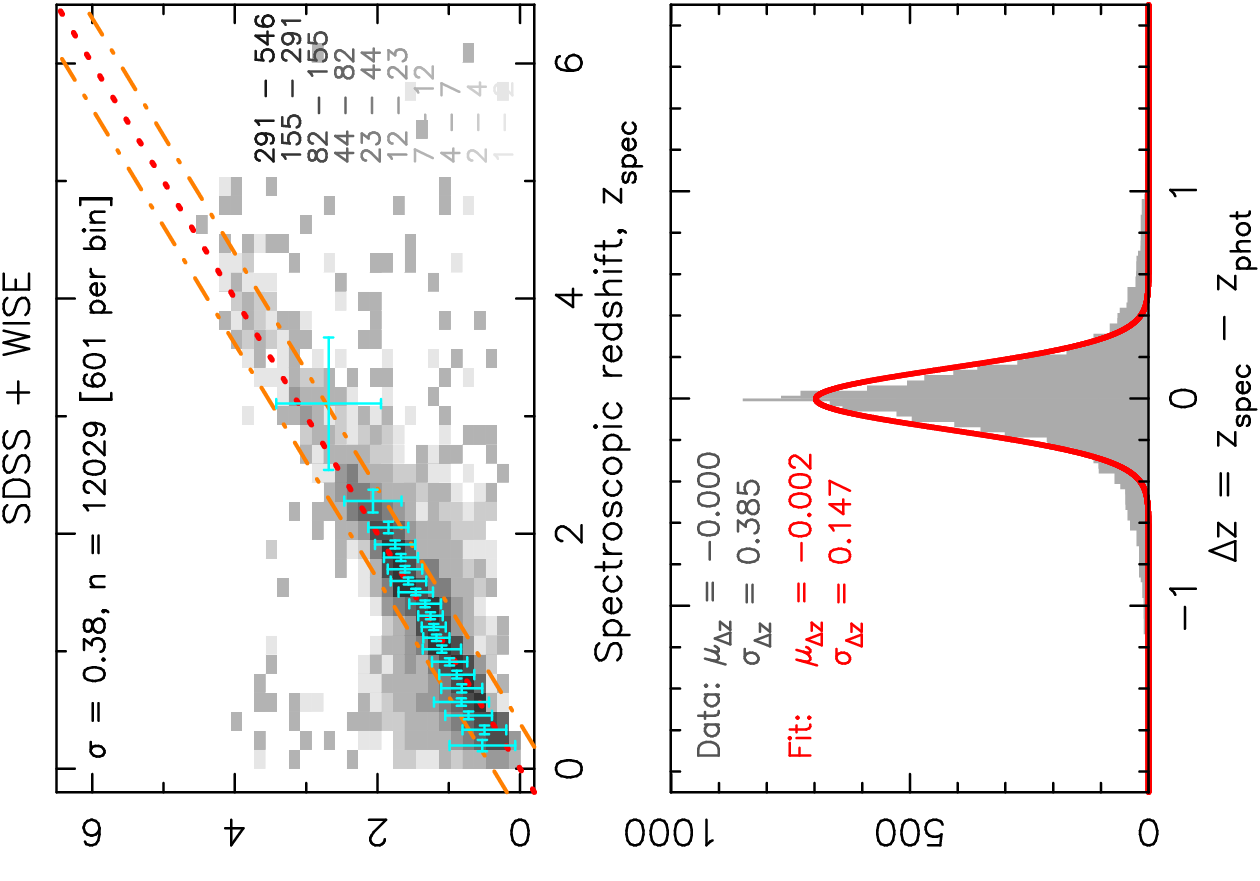}
\centering \includegraphics[angle=-90,scale=0.38]{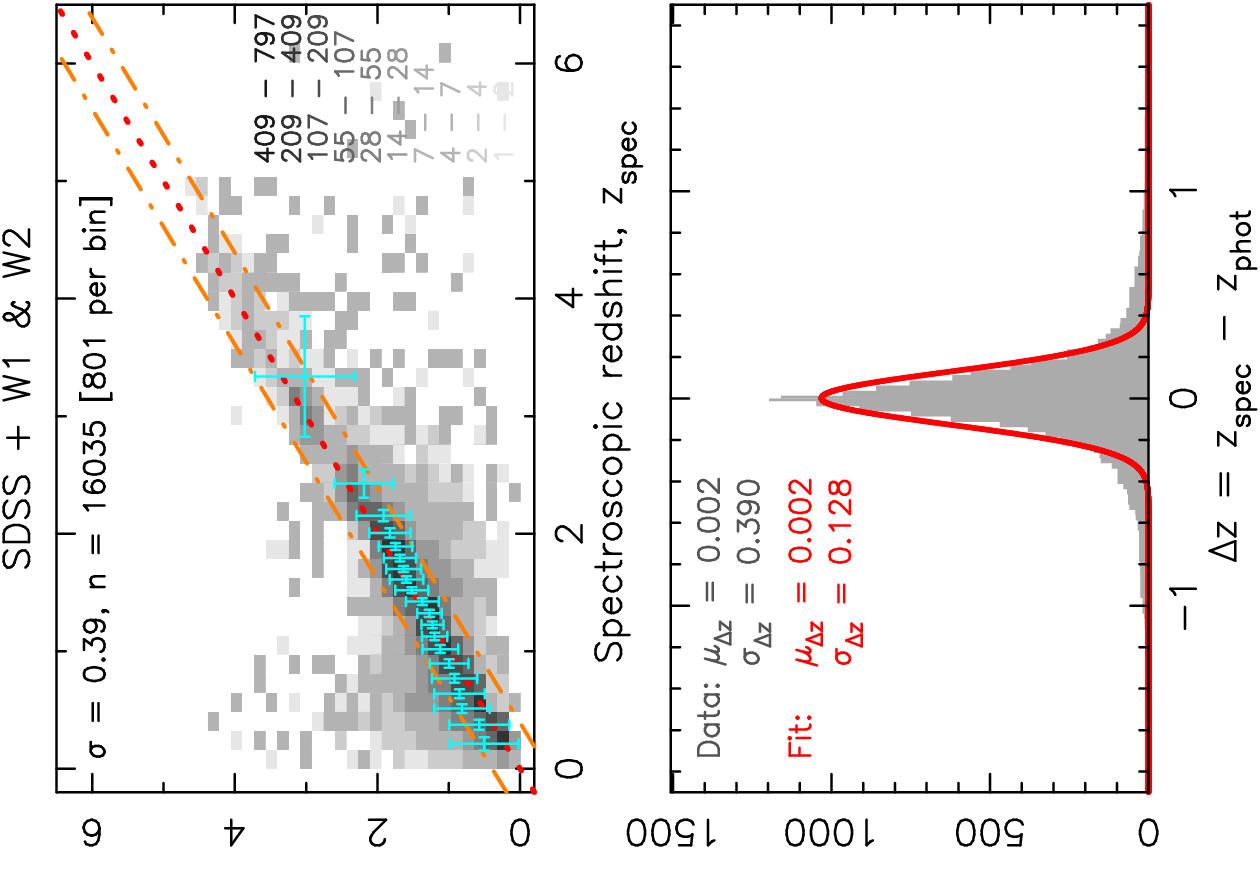}
\centering \includegraphics[angle=-90,scale=0.38]{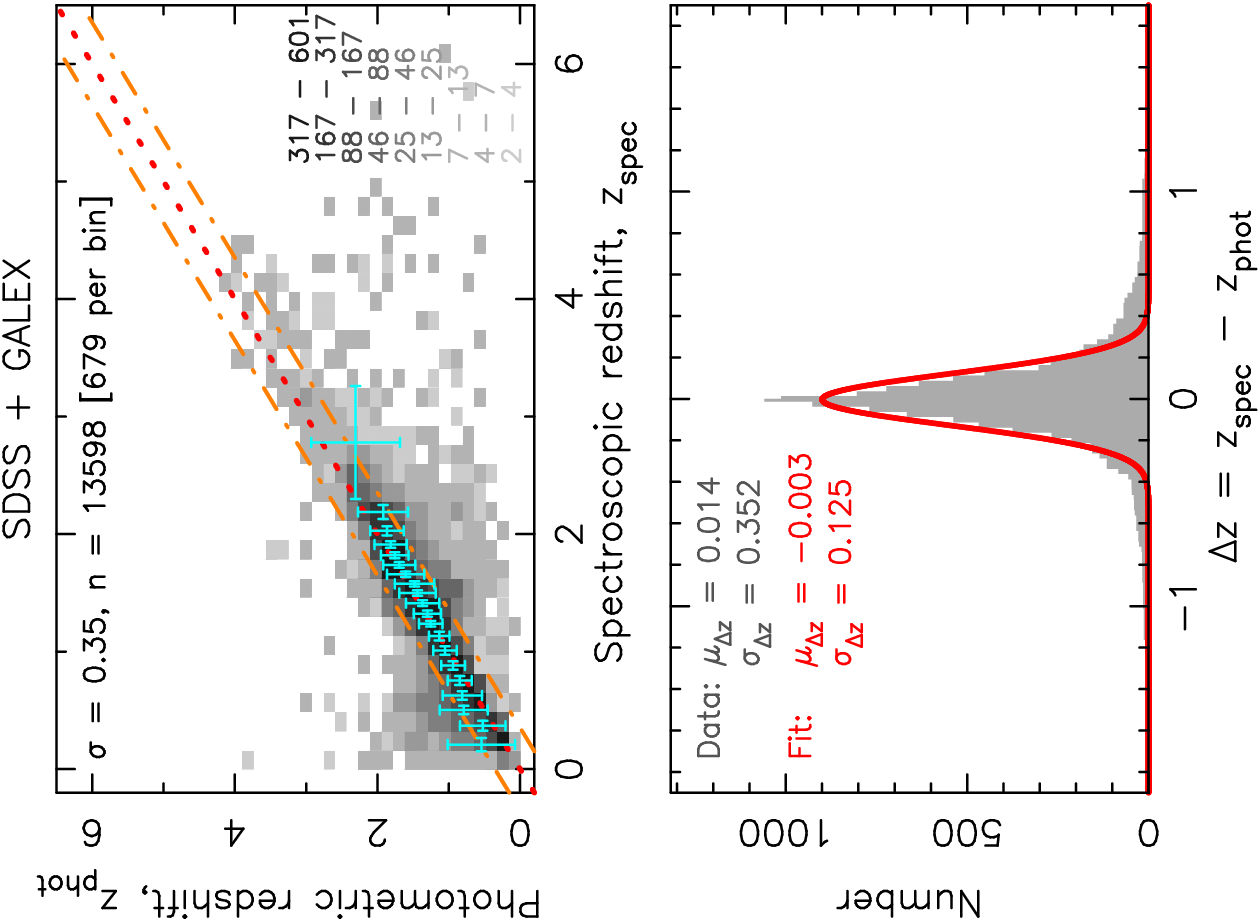}
\centering \includegraphics[angle=-90,scale=0.38]{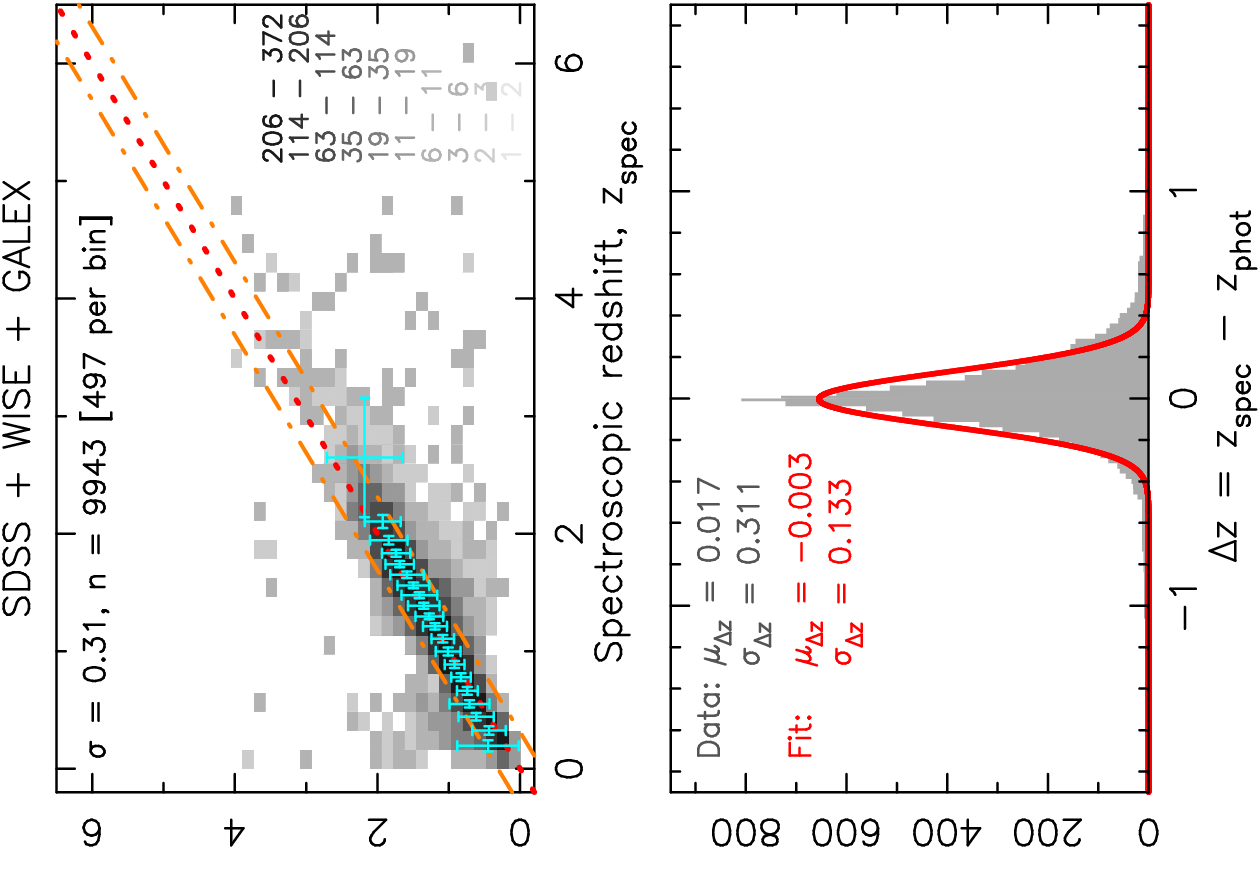}
\centering \includegraphics[angle=-90,scale=0.38]{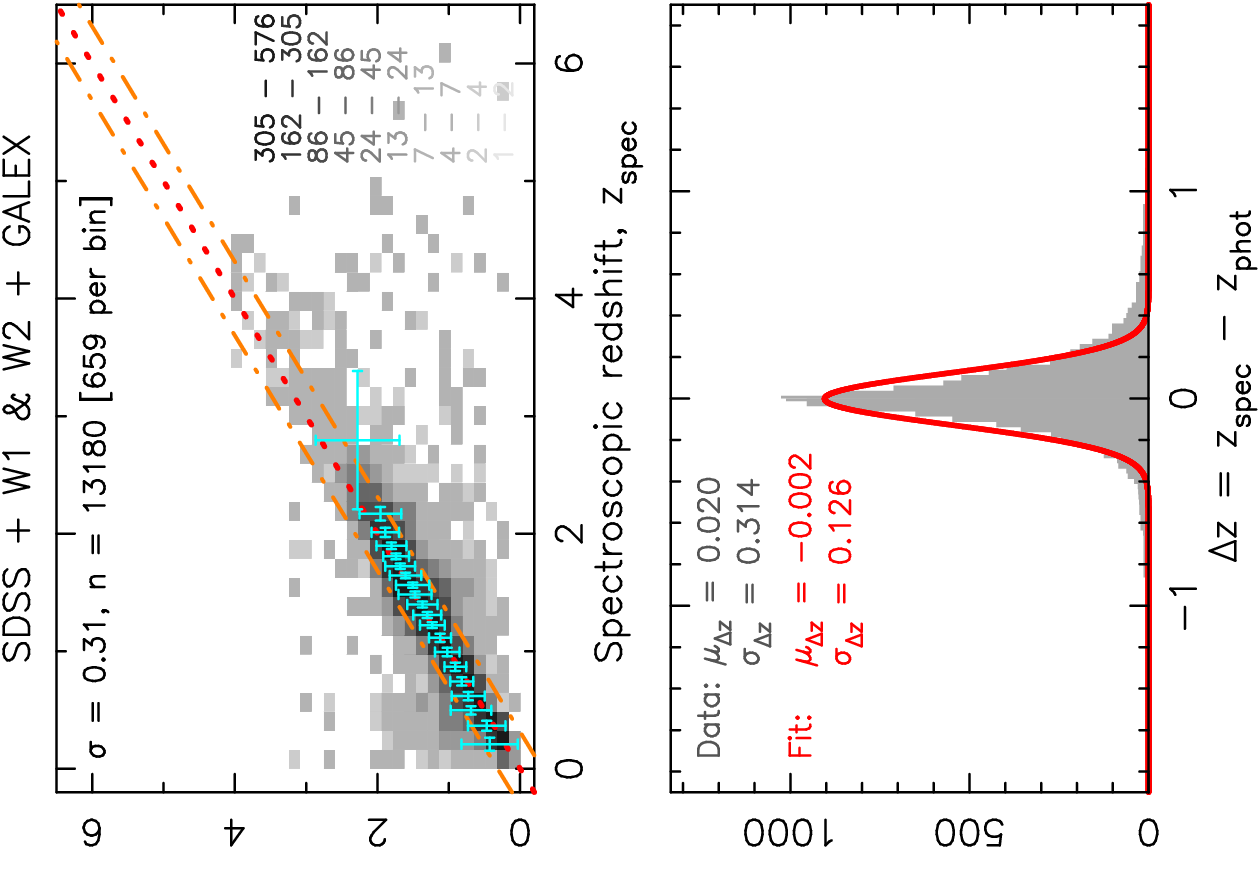}
\caption{The results from the various models using $k=10$ nearest neighbours. The grey-scale shows the photometric
  versus the spectroscopic redshift, with the key to the right showing the number of sources represented. The error bars
  show the mean values, with the $\pm1\sigma$ standard deviation in equally sized bins. The dotted line shows $z_{\rm phot} =
  z_{\rm spec}$, with  the dot--dashed lines showing the standard deviation  from this. The histogram below shows the
  $z_{\rm spec} - z_{\rm phot}$ distribution, where the mean and standard deviation is given for both the distribution
  and the Gaussian fit.  The outlier fractions are summarised in Table \ref{tab:SDSS}.}
\label{big}
\end{figure*}

We start by using the standard $u- g$, $g - r$, $r - i$ and $ i -  z$ colours \citep{rws+01,bbm+08},
including also the $r$ magnitude as a feature in the algorithm \citep{hdzz16}. Of the sample of 33\,643,
33\,166 have all five SDSS magnitudes, giving a 98\% matching coincidence. 
We trained the model on half of the sample, finding that  $k\approx10$ nearest neighbours minimised the 
standard deviation, $\sigma$,  from the $z_{\rm phot} = z_{\rm spec}$ line (Fig.~\ref{big}, top left). 

In the top panel of the figure, we see a similar distribution to those obtained by \citet{rws+01,wrs+04,hdzz16},
with two dense groups of outliers within $z_{\rm spec}\lapp2$ and $z_{\rm phot}\lapp2$. The
data also exhibit the deviation from $z_{\rm phot} = z_{\rm spec}$ at $z_{\rm spec}\gapp2$,
contributing to the wide wings in the $\Delta z$ distribution (\citealt{rws+01,wrs+04,mhp+12,prm+15,rmp+15,hdzz16,cm19}).

\subsubsection{WISE colours}
\label{sec:wise}

As stated in Sect.~\ref{intro}, given the dependence of the observed-frame bands on redshift, a single set of colours
would not be expected to yield accurate photometric redshifts over a wide range. Specifically, \citet{cm19} find that it
is the {\em rest-frame} $(U-K)/(W2-FUV)$ colour ratio which correlates strongly with redshift, corresponding to the {\em
  observed-frame} ratios $(I-W2)/(W3-U)$ at $1\leq z \leq3$ and $(I-W2.5)/(W4-R)$ at $z>3$.\footnote{\citeauthor{cm19}
refer to the $\lambda=8.0$~$\mu$m magnitude ({\em Spitzer Space Telescope}), located between $W2$ and $W3$,  as $W2.5$.}  They postulate that this
dependence
results from a decrease in the rest-frame $U-K$ colour coupled with an increase in the $W2 - FUV$ colour as the 
luminosity increases, which gives a proxy for the redshift via the Malmquist bias.

It is therefore apparent that in order to accurately determine a large range of photometric redshifts, other bands
beyond the optical must be invoked. The method of \citeauthor{cm19} requires nine different magnitude measures from a
number of disparate surveys, resulting in photometric redshifts being obtained for  $<34$\% of the sources.\footnote{The matching coincidence is 34\% for the observed-frame $W3\cap W2 \cap\, I\cap U$ magnitudes alone, which falls to 2\% for all
of the magnitudes required to cover the whole redshift range.} The WISE
database provides two near-infrared (NIR, $W1$ \& $W2$) and two mid-infrared (MIR, $W3$ \& $W4$) magnitudes from a
single survey and overlaps well with the sources in the SDSS database (e.g. \citealt{lhs16,slj+16,wwf+16}).
Adding the $z-W1$, $W1-W2$, $W2-W3$ and $W3-W4$ colours to the algorithm, we obtain  considerably better photometric
redshifts,
with  greatly reduced wings in the $\Delta z$ distribution (Fig.~\ref{big}, top middle). 
Using sources for which all four WISE magnitudes are available yields a 72\% matching coincidence, with the 
missing sources due to the objects being undetected in the $W3$ and  $W4$ bands.

The matching coincidence  can be increased to 95\%  by using the $W1$ and $W2$ magnitudes only\footnote{The 5\% 
of undetected sources are due to confusion in the NIR around the SDSS source.}, 
which returns a similar result to using all of the WISE magnitudes (Fig.~\ref{big}, top right).
This could be due to 
the $W3$ and $W4$ magnitudes probing a relatively featureless region of the SED (see Sect.~\ref{disc})\footnote{There are
silicate features at $\lambda = 10$ and 18~$\mu$m, but these cannot be detected at the coarse spectral resolutions
considered here.}, which does not
contribute significantly to the model.

\subsubsection{GALEX colours}

In addition to the infrared, the ultraviolet (UV) colours may also be used to improve upon the optical data alone:
\citet{bbm+08} combined the SDSS colours with the near-ultraviolet ($NUV$, $\lambda = 227$~nm) and far-ultraviolet
($FUV$, $\lambda = 153$~nm) bands of the {\em Galaxy Evolution Explorer} \citep{mfs+05} to obtain a standard deviation
of $\sigma = 0.34$ from 11\,149 QSOs. The GALEX database was also queried as part of the photometry search
(Sect.~\ref{sec:samp}) and, adding $FUV-NUV$ and $NUV-u$ to the SDSS colours in the algorithm (Fig.~\ref{big}, bottom
left), we obtain a similar standard deviation as \citeauthor{bbm+08}. This is a significant improvement over the SDSS
colours alone and a slight improvement over the SDSS+WISE colours.
However, due to the Lyman break (see Sect.~\ref{disc}), the signal in the GALEX data drops
significantly  at high redshift, particularly at $z\gapp3$ (see Fig.~\ref{SED}), 
which will limit the GALEX photometry's usefulness at high redshift.

Adding the WISE to the GALEX colours (Fig. \ref{big}, bottom middle), we see further improvement 
although at the cost of the matching coincidence falling to 59\%. 
As before, using the $W1$ and $W2$ magnitudes only gives a similar result while retaining a 78\% coincidence (Fig.~\ref{big}, bottom right).

\section{Discussion}
\label{disc}

By adding the WISE to the SDSS colours in the kNN algorithm, we obtain significantly better photometric
redshifts than from using the SDSS alone (cf. \citealt{rws+01,wrs+04,mhp+12,hdzz16}). Furthermore, this is 
without the need to filter out outliers (red sources, e.g. \citealt{rws+01}), nor the visual inspection of images prior
to  their inclusion in the algorithm (e.g. \citealt{mhp+12}). Even just the addition of the two NIR ($W1$ \&  $W2$)
magnitudes leads to significant improvement, which can be further enhanced with 
the addition of the GALEX magnitudes (Table~\ref{tab:SDSS}).
\begin{table}
\centering
\caption{Summary of the algorithm performances for the SDSS sample. $n$ gives the matching coincidence of sources (out of
  16\,850) and is followed by the percentage of photometric redshifts which lie within $\Delta z$ of the spectroscopic
  value.}
\begin{tabular}{@{}l  r  c  c c  c @{}} 
\hline
\smallskip
Algorithm & $n$ &  \multicolumn{3}{c}{Percentage within  $\Delta z$} \\
                   &           &  $\pm0.1$ &  $\pm0.2$   &  $\pm0.5$ \\
\hline
SDSS         &    16\,583 & 37.1 & 56.4 & 77.4 \\ 
SDSS + WISE  &  12\,029&  43.9 & 69.4 &  90.5 \\ 
SDSS + $W1$ \& $W2$  & 16\,035 & 47.7  & 71.3 &          90.2 \\  
SDSS + GALEX    & 13\,598 & 48.5  & 72.1 &  91.0 \\
SDSS + GALEX        & 13\,598 & 48.5  & 72.1 &  91.0 \\ 
SDSS + $W1$ \& $W2$ + GALEX & 13\,180 & 50.5 & 75.1 & 94.1 \\
\hline
\end{tabular}
\label{tab:SDSS}  
\end{table} 

\begin{figure}
\centering \includegraphics[angle=-90,scale=0.52]{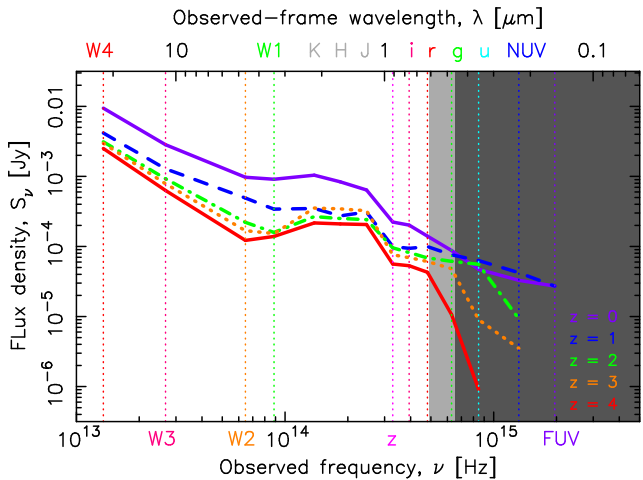}
\caption{Mean SEDs of the SDSS sample at $z> 0.2$ (29\,217 photometry points), $0.95 < z< 1.05$ (41\,383 points), $1.95
  < z< 2.05$ (33\,461 points), $2.8 < z< 3.2$ (16\,464 points) and $3.5 < z< 4.5$ (14\,464 points). The light grey
  region shows the range of the Lyman break due to $\lambda\leq1216$~\AA\ absorption and the dark region for
  $\lambda\leq912$~\AA\ absorption, at $z\leq4$.}
\label{SED}
\end{figure}

In Fig.~\ref{SED} we show the mean spectral energy distributions of the sample at various redshifts, where
 we see the  inflection at $\lambda\approx1$~$\mu$m, as the NIR emission from heated dust  
transitions to the optical emission from the accretion disk (\citealt{rob96} and references therein).
The excess at $\lambda\gapp1$~$\mu$m from the warm dust is also apparent (e.g. \citealt{bfl+19}), although only the
shape of the $JHK$ profile shows any redshift dependence. An increase in the peak frequency, which would counter the
redshift of the profile, would be expected as the luminosity (redshift) increases due to the higher peak temperature of a
modified blackbody (\citealt{cd19} and references therein).  The addition of the NIR bands permits the inclusion of
these features, explaining why the $K$ ($\lambda=2.2$~$\mu$m) and $W2$ bands are crucial to the empirical model
\citep{cm19}. Another feature is the Lyman break, where UV photons are absorbed by intervening hydrogen.  This is also
apparent in Fig.~\ref{SED}, particularly for the mean $z=4$ SED, where the first drop in flux at $\lambda_{\rm
  rest}=0.1216$~$\mu$m ($\lambda_{\rm obs}=0.61$~$\mu$m) is due to Lyman-\AL\ absorption and the second drop due to
ionisation by $\lambda_{\rm rest}=0.0912$~$\mu$m ($\lambda_{\rm obs}=0.46$~$\mu$m) photons. The addition of the GALEX
bands permits the inclusion of this feature at low redshift, although, 
the steep drop in flux results in a loss of signal at high redshift.  Thus, the incorporation of upper limits to the UV
fluxes into the algorithm would be required in order to fully utilise this feature.

As stated in Sect.~\ref{intro}, a photometric redshift estimate based only upon the source magnitudes will prove
invaluable to large surveys of redshifted continuum sources. Of particular interest are radio sources, in the era of the
{\em Square Kilometre Array} and its pathfinders. Given that, in practice, the detections from blind radio surveys 
will then be checked/re-observed for photometry in other bands, we wish to test the algorithm on an independent
radio selected sample which has been followed up for spectroscopic redshifts, rather than an optical sample
(SDSS), which has been cross-matched with radio sources.
We therefore use the {\em Optical Characteristics of
  Astrometric Radio Sources} (OCARS) catalogue of {\em Very Long Baseline Interferometry} astrometry sources, a sample
of flat spectrum radio sources (quasars) observed over five radio bands (spanning 2--27~GHz)\footnote{$S$-band
  (2--4~GHz), $C$-band (4--8~GHz), $X$-band (8--12~GHz), $U$-band (12--18~GHz) and $K$-band (18--27~GHz).}, with
$S$-band flux densities ranging from 15~mJy to 4.0~Jy \citep{mab+09}.  Of these, 3663 have spectroscopic redshifts
\citep{mal18}, with 36\% having the full SDSS photometry (Table~\ref{tab:OCARS}).\footnote{48\% of
  OCARS sources have at least one SDSS magnitude match, which is expected as the former covers the whole sky and the
  latter is restricted to northern declinations.} 
\begin{table}
\centering
  \caption{As Table~\ref{tab:SDSS}, but for the OCARS sample (out of 3663 sources).}
\begin{tabular}{@{}l  r  c  c c  c @{}} 
\hline
\smallskip
Algorithm & $n$ & \multicolumn{3}{c}{Percentage within  $\Delta z$} \\
                        &   &   $\pm0.1$ &  $\pm0.2$   &  $\pm0.5$ \\
\hline
SDSS   &          1320 & 27.1 & 44.4 & 66.9 \\
SDSS + WISE &  1007 & 37.3 & 56.2 & 80.6 \\
SDSS + $W1$ \& $W2$ & 1187 &    38.8 & 59.9 & 79.3 \\
SDSS + GALEX & 810 &  41.7 & 61.6 & 83.3 \\    
SDSS + WISE + GALEX    & 676 &     44.4 & 63.9 & 86.8 \\     
SDSS + $W1$ \& $W2$ + GALEX &739   &   47.9 & 68.9 & 89.3\\ 
\hline
\end{tabular}
\label{tab:OCARS}  
\end{table} 

\begin{figure*}
\setlength{\lineskip}{10pt}
\centering \includegraphics[angle=-90,scale=0.38]{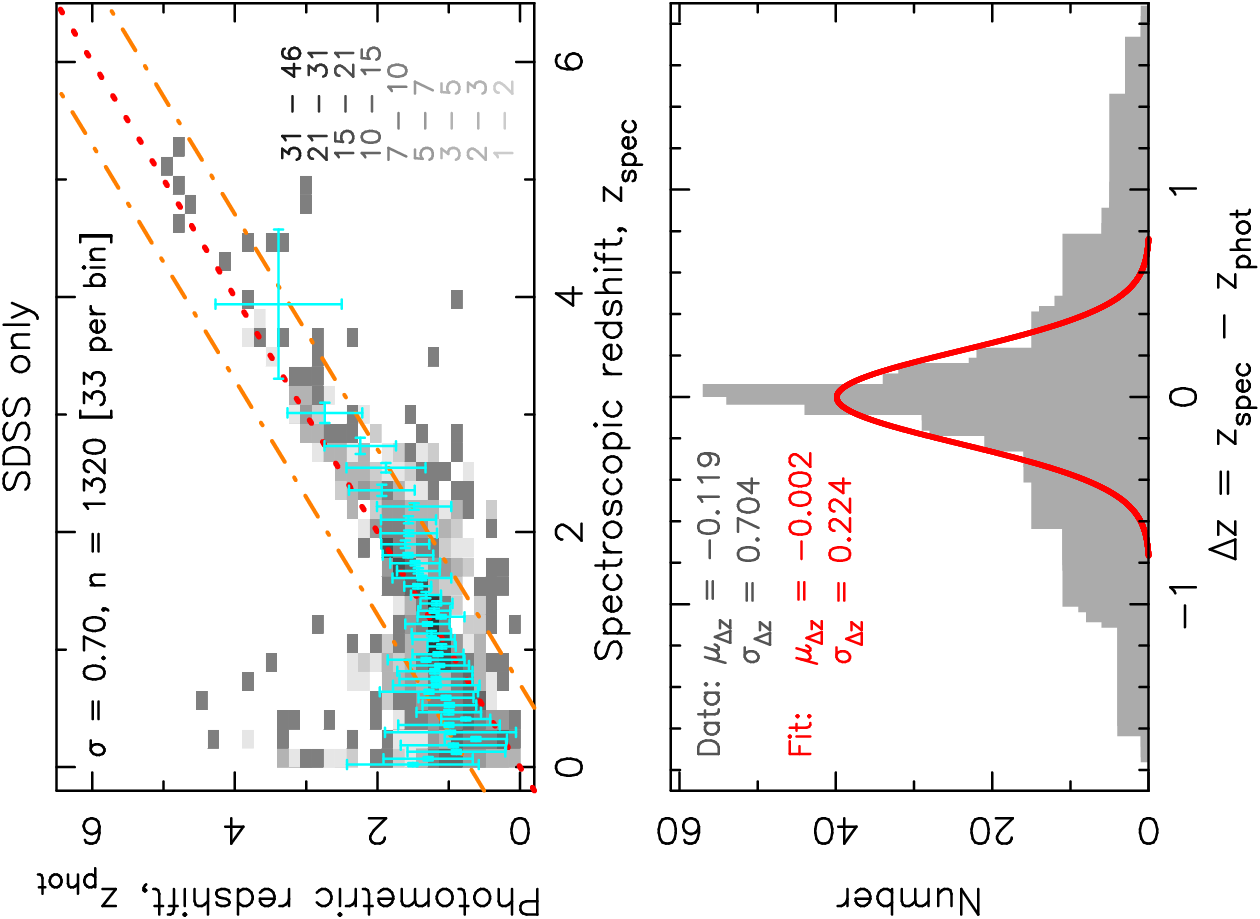}
\centering \includegraphics[angle=-90,scale=0.38]{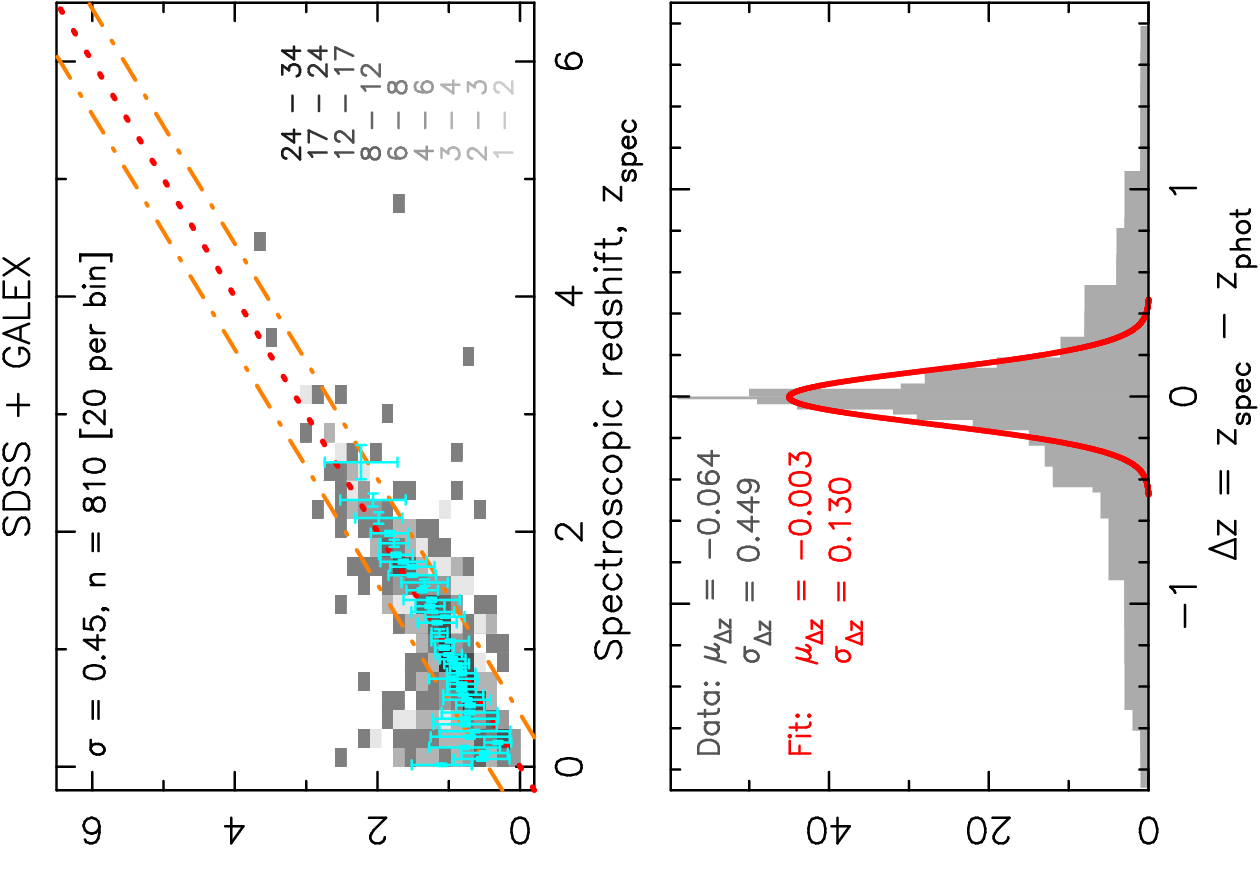}
\centering \includegraphics[angle=-90,scale=0.38]{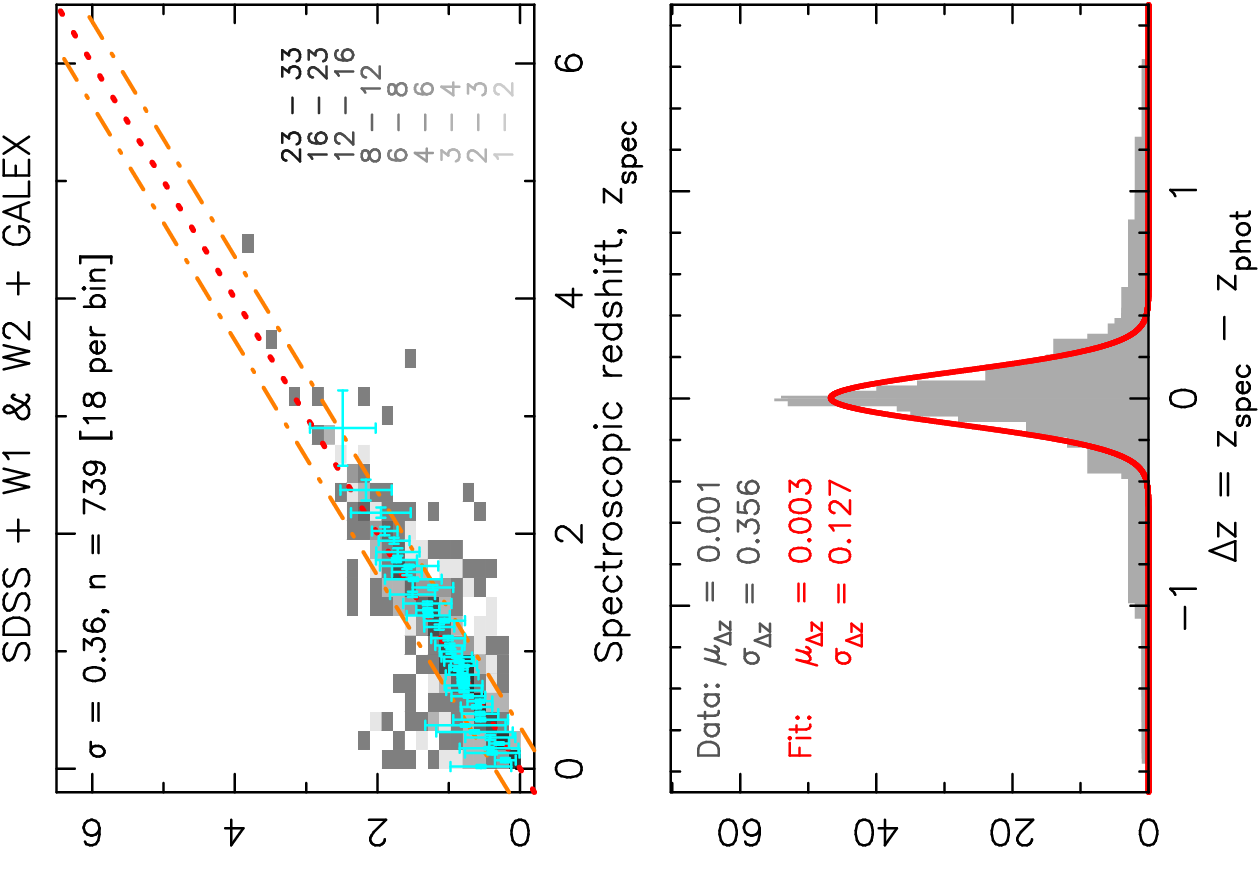}
\caption{As Fig.~\ref{big}, but using the SDSS sample to obtain the photometric redshifts of the OCARS sample.}
\label{OCARS_big}
\end{figure*}

Since our aim is to predict the photometric redshifts of a radio selected sample, with no a priori knowledge of the
spectroscopic redshifts, 
we train the algorithm on the full SDSS sample. The model is then applied to the OCARS sources with the relevant photometry,
as well as spectroscopic redshifts (in order to test the results). In Fig.~\ref{OCARS_big}, we see different distributions
than from the SDSS photometric redshifts (Fig.~\ref{big}),
\begin{figure}
\centering \includegraphics[angle=-90,scale=0.51]{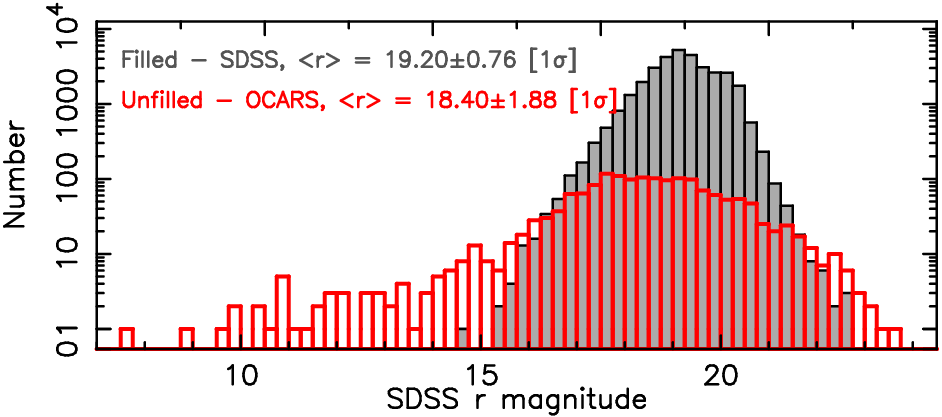}
\centering \includegraphics[angle=-90,scale=0.51]{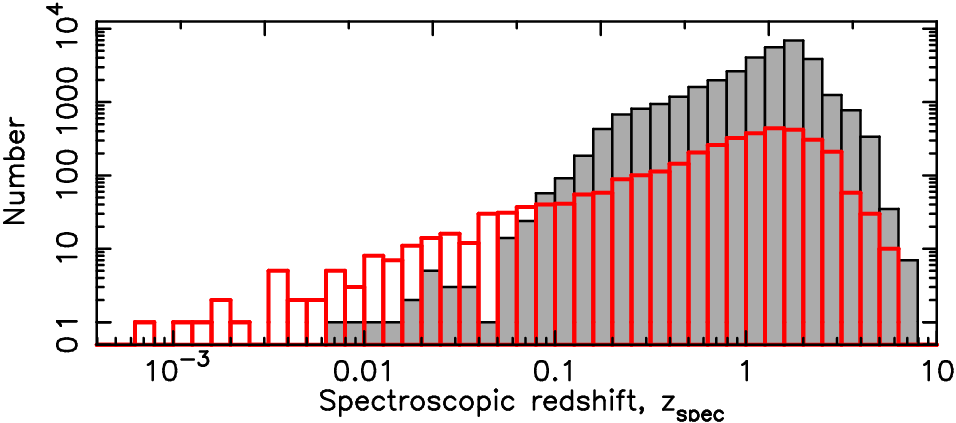}
\caption{The $r$ band magnitudes (top) and redshift distribution (bottom) for the SDSS (filled) and OCARS (unfilled histogram) sources.} 
\label{r-z}
\end{figure}
which is confirmed by the spread in magnitudes (Fig.~\ref{r-z}, top),
possibly due to the wider range of redshifts of the OCARS sample (Fig.~\ref{r-z}, bottom). Nevertheless, training the
algorithm on the SDSS sample gives photometric redshifts which are nearly as accurate as for the SDSS sources themselves
(Fig.~\ref{big}) when all three surveys are used (Fig.~\ref{OCARS_big}, right), with the percentages within $\Delta z$
just being slightly lower (Table~\ref{tab:OCARS}).

\section{Conclusions}

A rapid automated method of obtaining source redshifts would vastly increase the scientific value of large surveys of
radio continuum sources. One method which has shown much promise is the $k$-nearest neighbour algorithm which utilises
SDSS colours. For a sample of 33\,643 QSOs from the SDSS,  98\%
have all five magnitudes. Using these sources to train half of the data, we obtain a mean of $\mu_{\Delta z}\approx0$
for $z_{\rm phot} - z_{\rm spec}$ for the remaining half.  However, in common with other studies which apply the kNN
algorithm to the SDSS colours \citep{rws+01,wrs+04,mhp+12,hdzz16}, the $\Delta z$ distribution is wide-tailed with
$\sigma_{\Delta z}=0.52$.  Including the WISE colours in the algorithm significantly narrows the $\Delta z$ distribution
($\sigma_{\Delta
  z}=0.38$), 
which is further improved with the addition of the GALEX colours ($\sigma_{\Delta z}=0.31$). 
Without the need for visual inspection nor the filtering out of outliers, the spread is comparable to other studies (using
similar and differing methods) which include NIR photometry \citep{bbm+08,bmh+12,bcd+13,ywf+17,dbw+18,sih18}. 

Although inclusion of both the WISE and GALEX photometry significantly improves the photometric redshifts, we find that
exclusion of the WISE $W3$ \& $W4$ magnitudes has no effect, due to the FIR range of the SED being relatively
featureless. The $W1$ \& $W2$ magnitudes are, however, crucial since there is an excess in the NIR due
to emission  dust heated by the AGN.  Furthermore, the UV segment of the SED
exhibits the Lyman break, which becomes apparent in the optical band at redshifts of $z\gapp3$. Thus, the inclusion of
the GALEX photometry has the potential to distinguish high redshift sources, although the steep drop in signal limits
its usefulness.

For photometric redshifts obtained empirically from the observed source colours, \citet{cm19} found that different
combinations of observed-frame magnitudes were required for different redshift regimes. This method gave $\sigma_{\Delta
  z}=0.34$, although the numerous magnitude measurements required resulted in a low matching coincidence. Adding the
GALEX ultraviolet magnitudes to the SDSS and WISE colours in the kNN algorithm, we can surpass the empirical method
($\sigma_{\Delta z}=0.31$) while retaining a large matching coincidence.

Training the algorithm on the entire SDSS sample, the SDSS + $W1$ \& $W2$ + GALEX magnitude combination gives
$\mu_{\Delta z}=0.001$ and $\sigma_{\Delta z}=0.36$ when applied a sample of 3663 radio selected sources.  This, again,
markedly outperforms the algorithm trained on the SDSS colours alone ($\mu_{\Delta z}=-0.12$ and $\sigma_{\Delta
  z}=0.70$), although the matching coincidence is relatively low (20\%) due to the requirement of all three
surveys. This, however, is still significantly higher than that obtained from the empirical method (2\%,
\citealt{cm19}).\footnote{Which requires the $W4,W3,W2,K,I,R,U,FUV$ and 8~$\mu$m magnitudes.} Thus, we confirm that the addition of other wavebands, in particular the two WISE near-infrared ($W1$
\& $W2$) and two GALEX ultraviolet bands, can significantly improve the photometric redshifts obtained from the
$k$-nearest neighbour algorithm.

\section*{Acknowledgements}

I wish to thank the anonymous referee whose feedback helped significantly improve the manuscript.
This research has made use of the NASA/IPAC Extragalactic
Database (NED) which is operated by the Jet Propulsion Laboratory, California Institute of Technology, under contract
with the National Aeronautics and Space Administration and NASA's Astrophysics Data System Bibliographic Service. 
Funding for the SDSS has been provided by the Alfred P. Sloan Foundation, the Participating Institutions, the National Science Foundation, the U.S. Department of Energy, the National Aeronautics and Space Administration, the Japanese Monbukagakusho, the Max Planck Society, and the Higher Education Funding Council for England. 
 This publication makes use of data products from the Wide-field Infrared Survey
Explorer, which is a joint project of the University of California, Los Angeles, and the Jet Propulsion
Laboratory/California Institute of Technology, funded by the National Aeronautics and Space Administration.  This
publication makes use of data products from the Two Micron All Sky Survey, which is a joint project of the University of
Massachusetts and the Infrared Processing and Analysis Center/California Institute of Technology, funded by the National
Aeronautics and Space Administration and the National Science Foundation. GALEX is operated for NASA by the California
Institute of Technology under NASA contract NAS5-98034.  


\begin{thebibliography}{39}
\expandafter\ifx\csname natexlab\endcsname\relax\def\natexlab#1{#1}\fi

\bibitem[{{Alam} {et~al}\mbox{.}(2015){Alam}, {Albareti}, {Allende Prieto},
  {Anders}, {Anderson}, {Anderton}, {Andrews}, {Armengaud}, {Aubourg},
  {Bailey}, \& et~al.}]{aaa+15}
{Alam} S. {et~al.}, 2015, ApJS, 219, 12

\bibitem[{{Ananna} {et~al}\mbox{.}(2017){Ananna}, {Salvato}, {LaMassa}, {Urry},
  {Cappelluti}, {Cardamone}, {Civano}, {Farrah}, {Gilfanov}, {Glikman},
  {Hamilton}, {Kirkpatrick}, {Lanzuisi}, {Marchesi}, {Merloni}, {Nandra},
  {Natarajan}, {Richards}, \& {Timlin}}]{asu+17}
{Ananna} T.~T. {et~al.}, 2017, ApJ, 850, 66

\bibitem[{{Ball} {et~al}\mbox{.}(2008){Ball}, {Brunner}, {Myers}, {Strand},
  {Alberts}, \& {Tcheng}}]{bbm+08}
{Ball} N.~M., {Brunner} R.~J., {Myers} A.~D., {Strand} N.~E., {Alberts} S.~L.,
  {Tcheng} D., 2008, ApJ, 683, 12

\bibitem[{{Bianchini} {et~al}\mbox{.}(2019){Bianchini}, {Fabbian}, {Lapi},
  {Gonzalez-Nuevo}, {Gilli}, \& {Baccigalupi}}]{bfl+19}
{Bianchini} F., {Fabbian} G., {Lapi} A., {Gonzalez-Nuevo} J., {Gilli} R.,
  {Baccigalupi} C., 2019, ApJ, 871, 136

\bibitem[{{Bovy} {et~al}\mbox{.}(2012){Bovy}, {Myers}, {Hennawi}, {Hogg},
  {McMahon}, {Schiminovich}, {Sheldon}, {Brinkmann}, {Schneider}, \&
  {Weaver}}]{bmh+12}
{Bovy} J. {et~al.}, 2012, ApJ, 749, 41

\bibitem[{{Brescia} {et~al}\mbox{.}(2013){Brescia}, {Cavuoti}, {D'Abrusco},
  {Longo}, \& {Mercurio}}]{bcd+13}
{Brescia} M., {Cavuoti} S., {D'Abrusco} R., {Longo} G., {Mercurio} A., 2013,
  ApJ, 772, 140

\bibitem[{{Brown} {et~al}\mbox{.}(2014){Brown}, {Jarrett}, \& {Cluver}}]{bjc14}
{Brown} M.~J.~I., {Jarrett} T.~H., {Cluver} M.~E., 2014, PASA, 31, e049

\bibitem[{Curran \& Duchesne(2019)}]{cd19}
Curran S.~J., Duchesne S.~W., 2019, A\&A, 627, A93

\bibitem[{Curran {et~al}\mbox{.}(2019)Curran, {Hunstead}, {Johnston},
  {Whiting}, {Sadler}, {Allison}, \& {Athreya}}]{chj+19}
Curran S.~J., {Hunstead} R.~W., {Johnston} H.~M., {Whiting} M.~T., {Sadler}
  E.~M., {Allison} J.~R., {Athreya} R., 2019, MNRAS, 484, 1182

\bibitem[{Curran \& Moss(2019)}]{cm19}
Curran S.~J., Moss J.~P., 2019, A\&A, 629, A56

\bibitem[{Curran \& Whiting(2012)}]{cw12}
Curran S.~J., Whiting M.~T., 2012, ApJ, 759, 117

\bibitem[{{Curran} {et~al}\mbox{.}(2006){Curran}, {Whiting}, {Murphy}, {Webb},
  {Longmore}, {Pihlstr{\"o}m}, {Athreya}, \& {Blake}}]{cwm+06}
{Curran} S.~J., {Whiting} M.~T., {Murphy} M.~T., {Webb} J.~K., {Longmore}
  S.~N., {Pihlstr{\"o}m} Y.~M., {Athreya} R., {Blake} C., 2006, MNRAS, 371, 431

\bibitem[{{Duncan} {et~al}\mbox{.}(2018){Duncan}, {Brown}, {Williams}, {Best},
  {Buat}, {Burgarella}, {Jarvis}, {Ma{\l}ek}, {Oliver}, {R{\"o}ttgering}, \&
  {Smith}}]{dbw+18}
{Duncan} K.~J. {et~al.}, 2018, MNRAS, 473, 2655

\bibitem[{{Glowacki} {et~al}\mbox{.}(2017){Glowacki}, {Allison}, {Sadler},
  {Moss}, \& {Jarrett}}]{gas+18}
{Glowacki} M., {Allison} J.~R., {Sadler} E.~M., {Moss} V.~A., {Jarrett} T.~H.,
  2017, MNRAS, submitted (arXiv:1709.08634)

\bibitem[{{Han} {et~al}\mbox{.}(2016){Han}, {Ding}, {Zhang}, \&
  {Zhao}}]{hdzz16}
{Han} B., {Ding} H.-P., {Zhang} Y.-X., {Zhao} Y.-H., 2016, Research in
  Astronomy and Astrophysics, 16, 74

\bibitem[{{Johnston} {et~al}\mbox{.}(2008){Johnston}, {Taylor}, {Bailes},
  {Bartel}, {Baugh}, {Bietenholz}, {Blake}, {Braun}, {Brown}, {Chatterjee},
  {Darling}, {Deller}, {Dodson}, {Edwards}, {Ekers}, {Ellingsen}, {Feain},
  {Gaensler}, {Haverkorn}, {Hobbs}, {Hopkins}, {Jackson}, {James}, {Joncas},
  {Kaspi}, {Kilborn}, {Koribalski}, {Kothes}, {Landecker}, {Lenc}, {Lovell},
  {Macquart}, {Manchester}, {Matthews}, {McClure-Griffiths}, {Norris}, {Pen},
  {Phillips}, {Power}, {Protheroe}, {Sadler}, {Schmidt}, {Stairs},
  {Staveley-Smith}, {Stil}, {Tingay}, {Tzioumis}, {Walker}, {Wall}, \&
  {Wolleben}}]{jtb+08}
{Johnston} S. {et~al.}, 2008, Experimental Astronomy, 22, 151

\bibitem[{{Lang} {et~al}\mbox{.}(2016){Lang}, {Hogg}, \& {Schlegel}}]{lhs16}
{Lang} D., {Hogg} D.~W., {Schlegel} D.~J., 2016, AJ, 151, 36

\bibitem[{{Luken} {et~al}\mbox{.}(2019){Luken}, {Norris}, \& {Park}}]{lnp18}
{Luken} K.~J., {Norris} R.~P., {Park} L.~A.~F., 2019, PASP, 131, 108003

\bibitem[{{Ma} {et~al}\mbox{.}(2009){Ma}, {Arias}, {Bianco}, {Boboltz},
  {Bolotin}, {Charlot}, {Engelhardt}, {Fey}, {Gaume}, {Gontier}, {Heinkelmann},
  {Jacobs}, {Kurdubov}, {Lambert}, {Malkin}, {Nothnagel}, {Petrov},
  {Skurikhina}, {Sokolova}, {Souchay}, {Sovers}, {Tesmer}, {Titov}, {Wang},
  {Zharov}, {Barache}, {Boeckmann}, {Collioud}, {Gipson}, {Gordon}, {Lytvyn},
  {MacMillan}, \& {Ojha}}]{mab+09}
{Ma} C. {et~al.}, 2009, IERS Technical Note, 35, 1

\bibitem[{{Maddox} {et~al}\mbox{.}(2012){Maddox}, {Hewett}, {P{\'e}roux},
  {Nestor}, \& {Wisotzki}}]{mhp+12}
{Maddox} N., {Hewett} P.~C., {P{\'e}roux} C., {Nestor} D.~B., {Wisotzki} L.,
  2012, MNRAS, 424, 2876

\bibitem[{Majic \& Curran(2015)}]{maj15}
Majic R. A.~M., Curran S.~J., 2015, {Radio Photometric Redshifts: Estimating
  radio source redshifts from their spectral energy distributions}. Tech. rep.,
  Victoria University of Wellington

\bibitem[{{Malkin}(2018)}]{mal18}
{Malkin} Z., 2018, ApJS, 239, 20

\bibitem[{{Martin} {et~al}\mbox{.}(2005){Martin}, {Fanson}, {Schiminovich},
  {Morrissey}, {Friedman}, {Barlow}, {Conrow}, {Grange}, {Jelinsky},
  {Milliard}, {Siegmund}, {Bianchi}, {Byun}, {Donas}, {Forster}, {Heckman},
  {Lee}, {Madore}, {Malina}, {Neff}, {Rich}, {Small}, {Surber}, {Szalay},
  {Welsh}, \& {Wyder}}]{mfs+05}
{Martin} D.~C. {et~al.}, 2005, ApJ, 619, L1

\bibitem[{{Morganti} {et~al}\mbox{.}(2015){Morganti}, {Sadler}, \&
  {Curran}}]{msc+15}
{Morganti} R., {Sadler} E.~M., {Curran} S., 2015, Advancing Astrophysics with
  the Square Kilometre Array (AASKA14), 134

\bibitem[{{Norris} {et~al}\mbox{.}(2011){Norris}, {Hopkins}, {Afonso}, {Brown},
  {Condon}, {Dunne}, {Feain}, {Hollow}, {Jarvis}, {Johnston-Hollitt}, {Lenc},
  {Middelberg}, {Padovani}, {Prandoni}, {Rudnick}, {Seymour}, {Umana},
  {Andernach}, {Alexander}, {Appleton}, {Bacon}, {Banfield}, {Becker}, {Brown},
  {Ciliegi}, {Jackson}, {Eales}, {Edge}, {Gaensler}, {Giovannini}, {Hales},
  {Hancock}, {Huynh}, {Ibar}, {Ivison}, {Kennicutt}, {Kimball}, {Koekemoer},
  {Koribalski}, {L{\'o}pez-S{\'a}nchez}, {Mao}, {Murphy}, {Messias},
  {Pimbblet}, {Raccanelli}, {Randall}, {Reiprich}, {Roseboom},
  {R{\"o}ttgering}, {Saikia}, {Sharp}, {Slee}, {Smail}, {Thompson}, {Urquhart},
  {Wall}, \& {Zhao}}]{nha+11}
{Norris} R.~P. {et~al.}, 2011, PASA, 28, 215

\bibitem[{{Norris} {et~al}\mbox{.}(2019){Norris}, {Salvato}, {Longo},
  {Brescia}, {Budavari}, {Carliles}, {Cavuoti}, {Farrah}, {Geach}, {Luken},
  {Musaeva}, {Polsterer}, {Riccio}, {Seymour}, {Smol{\v c}i{\'c}}, {Vaccari},
  \& {Zinn}}]{nsl+19}
{Norris} R.~P. {et~al.}, 2019, PASP, 131, 108004

\bibitem[{{Peters} {et~al}\mbox{.}(2015){Peters}, {Richards}, {Myers},
  {Strauss}, {Schmidt}, {Ivezi{\'c}}, {Ross}, {MacLeod}, \& {Riegel}}]{prm+15}
{Peters} C.~M. {et~al.}, 2015, ApJ, 811, 95

\bibitem[{{Richards} {et~al}\mbox{.}(2015){Richards}, {Myers}, {Peters},
  {Krawczyk}, {Chase}, {Ross}, {Fan}, {Jiang}, {Lacy}, {McGreer}, {Trump}, \&
  {Riegel}}]{rmp+15}
{Richards} G.~T. {et~al.}, 2015, ApJS, 219, 39

\bibitem[{{Richards} {et~al}\mbox{.}(2001){Richards}, {Weinstein}, {Schneider},
  {Fan}, {Strauss}, {Vanden Berk}, {Annis}, {Burles}, {Laubacher}, {York},
  {Frieman}, {Johnston}, {Scranton}, {Gunn}, {Ivezi{\'c}}, {Nichol},
  {Budav{\'a}ri}, {Csabai}, {Szalay}, {Connolly}, {Szokoly}, {Bahcall},
  {Ben{\'{\i}}tez}, {Brinkmann}, {Brunner}, {Fukugita}, {Hall}, {Hennessy},
  {Knapp}, {Kunszt}, {Lamb}, {Munn}, {Newberg}, \& {Stoughton}}]{rws+01}
{Richards} G.~T. {et~al.}, 2001, AJ, 122, 1151

\bibitem[{Robson(1996)}]{rob96}
Robson I., 1996, Active Galactic Nuclei. John Wiley \& Sons, Chichester

\bibitem[{{Salim} {et~al}\mbox{.}(2016){Salim}, {Lee}, {Janowiecki}, {da
  Cunha}, {Dickinson}, {Boquien}, {Burgarella}, {Salzer}, \&
  {Charlot}}]{slj+16}
{Salim} S. {et~al.}, 2016, ApJS, 227, 2

\bibitem[{{Salvato}(2014)}]{sal14}
{Salvato} M., 2014, in IAU Symposium, Vol. 304, Multiwavelength AGN Surveys and
  Studies, {Mickaelian} A.~M., {Sanders} D.~B., eds., pp. 421--421

\bibitem[{{Salvato} {et~al}\mbox{.}(2019){Salvato}, {Ilbert}, \&
  {Hoyle}}]{sih18}
{Salvato} M., {Ilbert} O., {Hoyle} B., 2019, Nature Astronomy, 3, 212

\bibitem[{{Skrutskie} {et~al}\mbox{.}(2006){Skrutskie}, {Cutri}, {Stiening},
  {Weinberg}, {Schneider}, {Carpenter}, {Beichman}, {Capps}, {Chester},
  {Elias}, {Huchra}, {Liebert}, {Lonsdale}, {Monet}, {Price}, {Seitzer},
  {Jarrett}, {Kirkpatrick}, {Gizis}, {Howard}, {Evans}, {Fowler}, {Fullmer},
  {Hurt}, {Light}, {Kopan}, {Marsh}, {McCallon}, {Tam}, {Van Dyk}, \&
  {Wheelock}}]{scs+06}
{Skrutskie} M.~F. {et~al.}, 2006, AJ, 131, 1163

\bibitem[{{van Haarlem} {et~al}\mbox{.}(2013){van Haarlem}, {Wise}, {Gunst},
  {Heald}, {McKean}, {Hessels}, {de Bruyn}, {Nijboer}, {Swinbank}, {Fallows},
  {Brentjens}, {Nelles}, {Beck}, {Falcke}, {Fender}, {H{\"o}randel},
  {Koopmans}, {Mann}, {Miley}, {R{\"o}ttgering}, {Stappers}, {Wijers},
  {Zaroubi}, {van den Akker}, {Alexov}, {Anderson}, {Anderson}, {van Ardenne},
  {Arts}, {Asgekar}, {Avruch}, {Batejat}, {B{\"a}hren}, {Bell}, {Bell}, {van
  Bemmel}, {Bennema}, {Bentum}, {Bernardi}, {Best}, {B{\^i}rzan}, {Bonafede},
  {Boonstra}, {Braun}, {Bregman}, {Breitling}, {van de Brink}, {Broderick},
  {Broekema}, {Brouw}, {Br{\"u}ggen}, {Butcher}, {van Cappellen}, {Ciardi},
  {Coenen}, {Conway}, {Coolen}, {Corstanje}, {Damstra}, {Davies}, {Deller},
  {Dettmar}, {van Diepen}, {Dijkstra}, {Donker}, {Doorduin}, {Dromer}, {Drost},
  {van Duin}, {Eisl{\"o}ffel}, {van Enst}, {Ferrari}, {Frieswijk}, {Gankema},
  {Garrett}, {de Gasperin}, {Gerbers}, {de Geus}, {Grie{\ss}meier}, {Grit},
  {Gruppen}, {Hamaker}, {Hassall}, {Hoeft}, {Holties}, {Horneffer}, {van der
  Horst}, {van Houwelingen}, {Huijgen}, {Iacobelli}, {Intema}, {Jackson},
  {Jelic}, {de Jong}, {Juette}, {Kant}, {Karastergiou}, {Koers}, {Kollen},
  {Kondratiev}, {Kooistra}, {Koopman}, {Koster}, {Kuniyoshi}, {Kramer},
  {Kuper}, {Lambropoulos}, {Law}, {van Leeuwen}, {Lemaitre}, {Loose}, {Maat},
  {Macario}, {Markoff}, {Masters}, {McFadden}, {McKay-Bukowski}, {Meijering},
  {Meulman}, {Mevius}, {Middelberg}, {Millenaar}, {Miller-Jones}, {Mohan},
  {Mol}, {Morawietz}, {Morganti}, {Mulcahy}, {Mulder}, {Munk}, {Nieuwenhuis},
  {van Nieuwpoort}, {Noordam}, {Norden}, {Noutsos}, {Offringa}, {Olofsson},
  {Omar}, {Orr{\'u}}, {Overeem}, {Paas}, {Pandey-Pommier}, {Pandey}, {Pizzo},
  {Polatidis}, {Rafferty}, {Rawlings}, {Reich}, {de Reijer}, {Reitsma},
  {Renting}, {Riemers}, {Rol}, {Romein}, {Roosjen}, {Ruiter}, {Scaife}, {van
  der Schaaf}, {Scheers}, {Schellart}, {Schoenmakers}, {Schoonderbeek},
  {Serylak}, {Shulevski}, {Sluman}, {Smirnov}, {Sobey}, {Spreeuw}, {Steinmetz},
  {Sterks}, {Stiepel}, {Stuurwold}, {Tagger}, {Tang}, {Tasse}, {Thomas},
  {Thoudam}, {Toribio}, {van der Tol}, {Usov}, {van Veelen}, {van der Veen},
  {ter Veen}, {Verbiest}, {Vermeulen}, {Vermaas}, {Vocks}, {Vogt}, {de Vos},
  {van der Wal}, {van Weeren}, {Weggemans}, {Weltevrede}, {White}, {Wijnholds},
  {Wilhelmsson}, {Wucknitz}, {Yatawatta}, {Zarka}, {Zensus}, \& {van
  Zwieten}}]{vwg+13}
{van Haarlem} M.~P. {et~al.}, 2013, A\&A, 556, A2

\bibitem[{{Wang} {et~al}\mbox{.}(2016){Wang}, {Wu}, {Fan}, {Yang}, {Yi},
  {Bian}, {McGreer}, {Yang}, {Ai}, {Dong}, {Zuo}, {Jiang}, {Green}, {Wang},
  {Cai}, {Wang}, \& {Yue}}]{wwf+16}
{Wang} F. {et~al.}, 2016, ApJ, 819, 24

\bibitem[{{Weinstein} {et~al}\mbox{.}(2004){Weinstein}, {Richards},
  {Schneider}, {Younger}, {Strauss}, {Hall}, {Budav{\'a}ri}, {Gunn}, {York}, \&
  {Brinkmann}}]{wrs+04}
{Weinstein} M.~A. {et~al.}, 2004, ApJS, 155, 243

\bibitem[{{Wright} {et~al}\mbox{.}(2010){Wright}, {Eisenhardt}, {Mainzer},
  {Ressler}, {Cutri}, {Jarrett}, {Kirkpatrick}, {Padgett}, {McMillan},
  {Skrutskie}, {Stanford}, {Cohen}, {Walker}, {Mather}, {Leisawitz}, {Gautier},
  {McLean}, {Benford}, {Lonsdale}, {Blain}, {Mendez}, {Irace}, {Duval}, {Liu},
  {Royer}, {Heinrichsen}, {Howard}, {Shannon}, {Kendall}, {Walsh}, {Larsen},
  {Cardon}, {Schick}, {Schwalm}, {Abid}, {Fabinsky}, {Naes}, \&
  {Tsai}}]{wem+10}
{Wright} E.~L. {et~al.}, 2010, AJ, 140, 1868

\bibitem[{{Yang} {et~al}\mbox{.}(2017){Yang}, {Wu}, {Fan}, {Jiang}, {McGreer},
  {Green}, {Yang}, {Schindler}, {Wang}, {Zuo}, \& {Fu}}]{ywf+17}
{Yang} Q. {et~al.}, 2017, AJ, 154

\end{thebibliography}

\label{lastpage}

\end{document}